\documentclass[11pt,preprint]{aastex}

\def\t0{\theta_{\circ}}

\def\be{\begin{equation}}
\def\en{\end{equation}}

\def\lsun{L_{\sun}}

\def\mdot{\dot{M}}

\def\h2{H$_2$}

\begin{document}

\title
{Far-Ultraviolet \h2 Emission from Circumstellar Disks}

\author{Laura Ingleby\altaffilmark{1}, Nuria Calvet\altaffilmark{1}, Edwin Bergin\altaffilmark{1}, Ashwin Yerasi\altaffilmark{1}, Catherine Espaillat\altaffilmark{1}, Gregory Herczeg\altaffilmark{2},  Evelyne Roueff\altaffilmark{3}, Herv{\'e} Abgrall\altaffilmark{3}, Jesus Hern{\'a}ndez\altaffilmark{4}, C{\'e}sar Brice{\~n}o\altaffilmark{4},  Ilaria Pascucci\altaffilmark{5}, Jon Miller\altaffilmark{1}, Jeffrey Fogel\altaffilmark{1},  Lee Hartmann\altaffilmark{1},  Michael Meyer\altaffilmark{6}, John Carpenter\altaffilmark{7}, Nathan Crockett\altaffilmark{1}, Melissa McClure\altaffilmark{1}
}

\altaffiltext{1}{Department of Astronomy, University of Michigan, 830 Dennison Building, 500 Church Street, Ann Arbor, MI 48109; lingleby@umich.edu, ncalvet@umich.edu, ebergin@umich.edu, ashwinry@umich.edu, ccespa@umich.edu, fogel@umich.edu, lhartm@umich.edu, jonmm@umich.edu, ncrocket@umich.edu, melisma@umich.edu}
\altaffiltext{2}{Max-Planck-Institut fur extraterrestriche Physik, Postfach 1312, 85741 Garching, Germany; gregoryh@mpe.mpg.de}
\altaffiltext{3}{LUTH and UMR 8102 du CNRS, Observatoire de Paris, Section de Meudon, Place J. Janssen, 92195 Meudon, France; evelyne.roueff@obspm.fr, herve.abgrall@obspm.fr}
\altaffiltext{4}{Centro de Investigaciones de Astronom{\'i}a (CIDA), M{\'e}rida 5101-A, Venezuela; jesush@cida.ve, briceno@cida.ve}
\altaffiltext{5}{Department of Physics and Astronomy, Johns Hopkins University, Baltimore, MD 21218; pascucci@pha.jhu.edu }
\altaffiltext{6}{ETH Hoenggerberg Campus, Physics Department, CH-8093 Zurich, Switzerland; mmeyer@phys.ethz.ch}
\altaffiltext{7}{Department of Astronomy, California Institute of Technology, Mail Code 249-17, 1200 East California Boulevard, Pasadena, CA 91125; jmc@astro.caltech.edu}

\begin{abstract}
We analyze the far-ultraviolet (FUV) spectra of 33 classical T Tauri
stars (CTTS), including 20 new spectra obtained with the Advanced
Camera for Surveys Solar Blind Channel (ACS/SBC) on the Hubble Space
Telescope.  Of the sources, 28 are in the $\sim$1 Myr old
Taurus-Auriga complex or Orion Molecular Cloud, 4 in the 8-10 Myr old
Orion OB1a complex and one, TW Hya, in the 10 Myr old TW Hydrae
Association.  We also obtained FUV ACS/SBC spectra of 10 non-accreting
sources surrounded by debris disks with ages between 10 and 125 Myr.
We use a feature in the FUV spectra
due mostly to electron impact excitation of
\h2 to 
study the evolution of the gas in the inner disk.  We find that the \h2 feature is absent in
non-accreting sources, but is detected in the spectra of CTTS and
correlates with accretion luminosity.  Since all young stars have
active chromospheres which produce strong X-ray and UV emission
capable of exciting \h2 in the disk, the fact that the non-accreting
sources show no \h2 emission implies that the \h2 gas in the inner
disk has dissipated in the non-accreting sources, although dust (and
possibly gas) remains at larger radii.  Using the flux at 1600 {\AA},
we estimate that the column density of \h2 left in the inner regions of the debris disks in our sample is less than $\sim 3\times10^{-6}\;{\rm g}$ ${\rm
cm^{-2}}$, nine orders of magnitude below the surface density of the
minimum mass solar nebula at 1 AU.

\end{abstract}


\section{ Introduction} 

Gas comprises 99\% of the mass of primordial disks.  As time increases, it is accreted onto the star,
formed into planets, and lost by photoevaporation, leaving
behind a debris disk, in which most of the mass is locked into planets
and other solid bodies traced by secondary dust arising
from collisions.  Although the general outline of this
process is agreed upon, many specific questions remain
unanswered, mainly because the gas is difficult to observe.  As a result, only $\sim$1\% of  the disk mass, the dust, has been used as a probe of
the disk evolution.  However, although interconnected, the evolution of
gas and dust may take different paths \citep{pascucci09}, making observations of the gas itself necessary to understand these processes. Of
particular importance are observations of the gas in the inner disk,
because it sets the chemical and physical conditions for planet
formation.  The bulk of the gas in these cold disks is in \h2, which lacks a permanent dipole component, so the pure
rotational and rovibrational lines are weak. Nonetheless, extensive
surveys of these lines in primordial disks have been carried out
(Bary et al. 2008; Bitner et al. 2008, and references therein), and
they have been detected in a handful of objects. Searches using less abundant molecules have also succeeded and provided
information on the gas in the inner region of gas-rich
disks \citep{carr08,salyk08,pascucci09,najita08}.  
Gas  has also been searched for in disks of more evolved sources which are no longer accreting, within the age range when the
transition from primordial to debris is supposed to happen, 
$\sim $5 - 20 Myr.  In particular, \cite{pascucci06} looked for \h2 in the
disks of several non-accreting sources and found that
the amount of gas still present at 5 - 20 Myr is not large enough to
form the gas giant planets at that time.  This observation agrees with
results indicating that the amount of hot gas in disks of
non-accreting sources is decreased when compared to accreting sources
\citep{carmona07}.
 
UV observations are very promising for detecting the gas. The strong
stellar Ly$\alpha$ radiation bathes the UV thin regions of the circumstellar
material and, as long as the \h2 has a temperature of a few thousand
degrees, the line excites electrons to upper electronic states,
which produces a plethora of emission lines in the UV when they
de-excite 
\citep[][H06, and references therein]{herczeg06}.
At the
same time, the
stellar high energy radiation fields eject electrons from heavy metals, and the
resulting free electrons produce additional electrons by
ionizing H and He atoms; these secondary electrons then excite \h2 to
upper levels, resulting in a characteristic spectrum of lines and
continuum in the UV 
\citep[][B0]{spitzer68,bergin04}.
For electron excitation to work efficiently, temperatures need to be high enough for neutral H to be present. The
relatively high temperature requirements mean that the \h2 detected
by these means must either to be close to the star or to be excited by
shocks. UV \h2 emission has been
found to be extended in objects surrounded by substantial natal
material, in the regions where the stellar outflow shocks this
material, or in fast accretors, where the \h2 may arise in the 
high density outflow
itself (H06).  However, without remnant envelopes such as the objects in this study, the only known exception being T Tau, the most likely place to find the required
high temperatures is in the inner disk. This makes the UV \h2
emission ideal for probing the \h2 gas in the innermost regions of 
disks, regions which are difficult to access by other means.

We
obtained ACS/SBC prism spectra of a fair number of accreting
Classical T Tauri stars (CTTS), non-accreting weak T Tauri stars
(WTTS), and more evolved disks (DD), covering the interesting age range,
$\sim$1 - 100 Myr. Our goal was to search for UV \h2 emission and
study its evolution. The poor spectral resolution of the ACS spectra
made the identification of Ly$\alpha$ fluorescent lines
impossible.  However, we were able to identify a feature around
$\sim1600$ {\AA}, first proposed by B04 as due mostly to electron impact
excitation of \h2. In this letter we present and analyze these
spectra.  We show that the \h2 feature is absent in all non-accreting
and evolved stars while present in all accreting stars, and use UV
fluxes to give very rough estimates of upper limits for the remaining surface
density of \h2 in the latter.

\section{Observations}
\label{obs}
We obtained observations of 20 CTTS and 10 non-accreting and evolved targets using
the Advanced Camera for Surveys Solar Blind Channel (ACS/SBC) on the
Hubble Space Telescope in 2007.  The observations were obtained in GO
programs 10810 (PI: Bergin), 10840 (PI: Calvet) and 11199 (PI:
Hartmann).  Each ACS observation consists of a brief image in the
F165LP filter and a longer image obtained with the PR130L prism.  Images appear unresolved.  Offsets between the target location in the filter and prism image,
including the wavelength solution, were obtained from \cite{larsen06}.
The target spectrum was then extracted from a 41-pixel (1.3'') wide
extraction window.  Background count rates of 0.05 - 0.1 counts~s$^{-1}$
were calculated from offset windows and subtracted from the extracted
spectrum.  The absolute wavelength solution was then determined by
fitting the bright \ion{C}{4} $\lambda1549$ \AA\ doublet.  Fluxes were
calibrated from the sensitivity function obtained from white dwarf
standard stars by \cite{bohlin07}.  The spectra range from
1230--1900~{\AA} with a 2-pixel resolution of $\sim 300$ at 1230~\AA\
and $\sim 80$ at 1600~\AA.

Table \ref{tabprop} lists the ACS targets used in this analysis and
the properties of these objects.  The CTTS sources include 16 objects
in the Taurus-Auriga molecular cloud and four sources in the 25 Ori
aggregate in the Orion OB1a subassociation.  Spectral types for the
CTTS in Taurus are from \cite{furlan06}, and ages from \cite{hartmann03}.  To correct for reddening we used the law towards the star HD 29647
\citep{whittet04} and estimated $A_V$ by de-reddening the median
photometry of \cite{herbst94} to fit the fluxes of a standard star in
the region of the spectrum (V to J bands) where the emission is mostly
photospheric\footnote{Targets with high mass accretion rate,
as DL Tau and DR Tau show significant veiling at J \citep{edwards06},
so the estimated extinction may be in error, although it is consistent
with values from Taurus.}. 
We obtained accretion luminosities $L_{acc}$ for the Taurus sources
using the U band excesses following \citet{gullbring98}, 
and the median U from photometry in \cite{herbst94}.
The ages, spectral
types, luminosities, $A_V$'s, and $L_{acc}$ for the sources in 25 Ori were taken from \cite{briceno07,hernandez07} and \cite{calvet05}.

\begin{figure}
\plotone{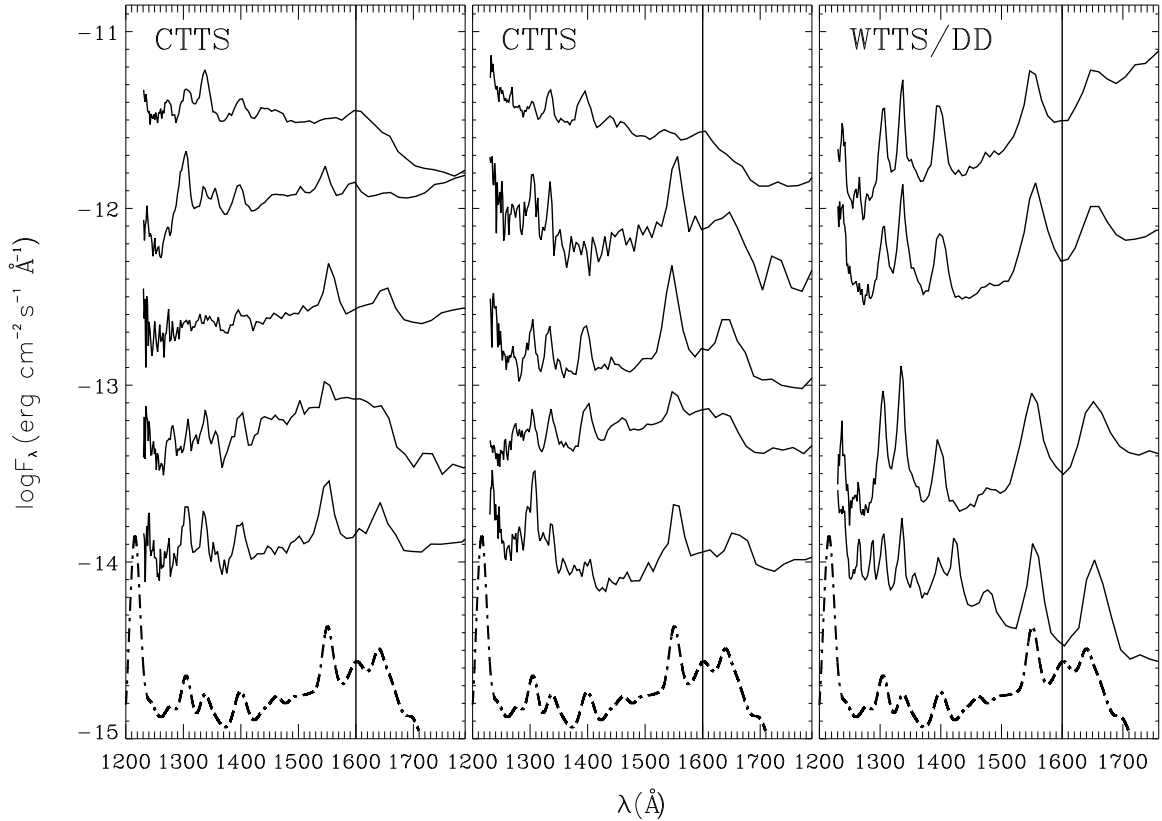}
\caption{Sample of ACS CTTS spectra.  Spectra have been corrected for reddening using the values of $A_V$ listed in Table \ref{tabprop}.  Spectra have been scaled vertically for clarity.  The bottom spectrum (dash-dotted line) in each panel is the STIS TW Hya spectrum smoothed to the resolution of the ACS spectra for comparison and offset by -1.2.  The vertical line at 1600 {\AA} marks the center of the feature used to identify the \h2. Left panel, from top to bottom including the offset in parenthesis: DP Tau (+2.7), DR Tau (+1.5), FM Tau (+0.5), FP Tau (+2.2) and GK Tau (+0.3).  Middle panel; from top to bottom: HN Tau A (+1.9), HN Tau B (+3.0), IP Tau (+1.5), UZ Tau A (+1.2) and  UZ Tau B (+0.45).  The right panel shows ACS spectra of WTTS/DD; from top to bottom: HD 12039 (+3.4), HD 202917 (+2.5), HD 61005 (+2.0), HD 92945 (+1.7) and HD 98800 (+0.8).
}
\label{spectra}
\end{figure}

The non-accreting sources (WTTS/DD) were selected to have no evidence
of accretion and to have excesses in either Spitzer Space Telescope Infrared Spectrograph (IRS) spectra or 24 and 70
$\mu$m Multiband Imaging Photometer (MIPS) photometry, indicating the presence of debris disks.
The sources in the TW Hydrae Association have been identified as
WTTS by spectral observations which showed
H$\alpha$ in emission \citep{webb99} and strong Li 6707 in absorption
\citep{kastner97}.  The WTTS/DD and their
properties were discussed in \cite{carpenter09,carpenter08},
\cite{hillenbrand08}, \cite{verrier08}, \cite{chen05} and
\cite{low05}.  Examples of the ACS target spectra are shown in Figure
\ref{spectra}.

We supplemented the ACS data with previously published medium and high
resolution STIS data of CTTS \citep[][B04]{calvet04,herczeg02,herczeg04}.  The
source properties, listed in Table \ref{tabprop}, were taken from
\cite{calvet04} for the Orion Molecular Cloud sources, and derived 
as described for the ACS Taurus sources for the STIS Taurus sources.
We adopt the spectral type and age from \cite{webb99} and $A_V$
from \cite{herczeg04} for TW Hya.  Accretion luminosities for the
STIS sample were taken from \cite{calvet04} and Ingleby et al. (2009).

\section{Results}
\label{analysis}

\begin{figure}
\plotone{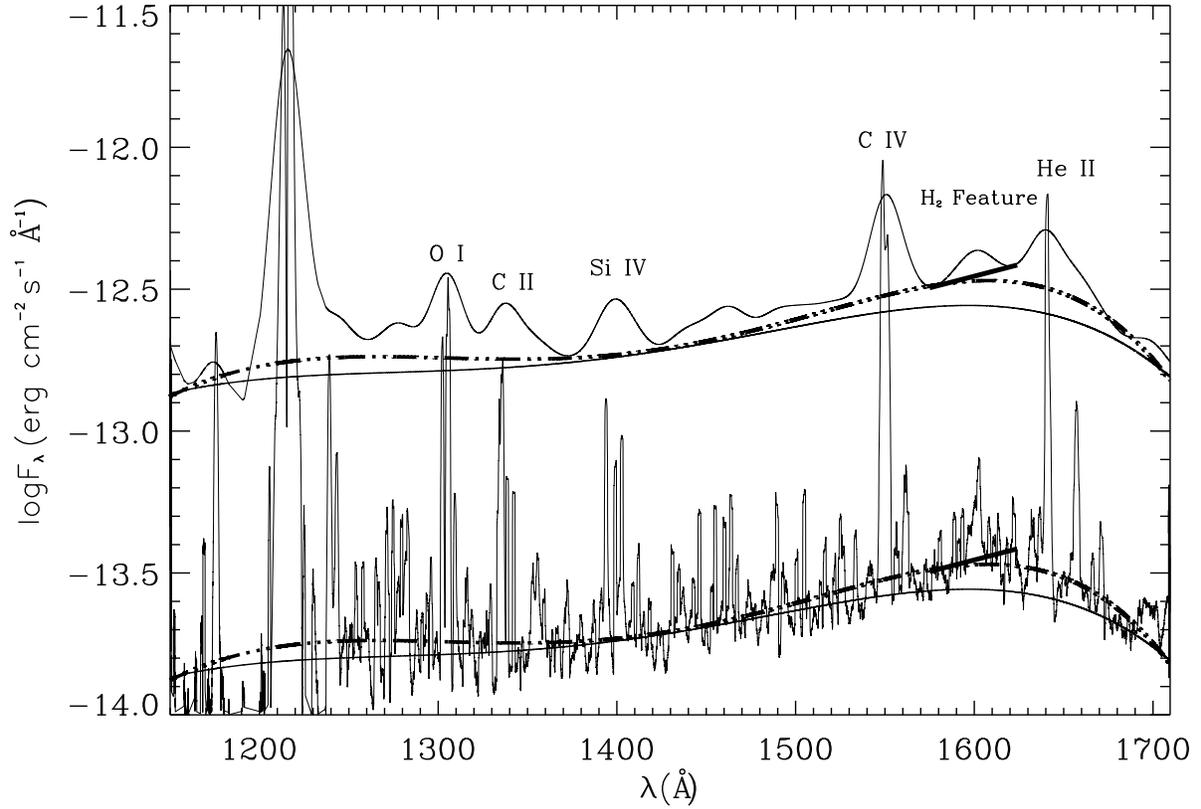}
\caption{Observed and convolved spectra for TW Hya.  The bottom spectrum is the high resolution STIS FUV spectrum.  The top spectrum is the TW Hya spectrum convolved to the ACS spectral resolution and offset by +1.0.  The solid and dashed lines on the smoothed spectrum show the three subtracted continua.  These three continua are also shown plotted on the high resolution spectrum and indicate that the lowest continuum may provide the best measure of the luminosity.  The strong emission lines are labeled along with the \h2 feature.
}
\label{smtw}
\end{figure}

Following B04, we identified a feature in the STIS spectra
at 1600 {\AA} which is due mostly to electron impact \h2 emission.  Due to the low resolution of the ACS spectra,
we used the high resolution spectrum of TW Hya \citep{herczeg04} to identify this
feature in the ACS spectra;  in
Figure \ref{smtw} we compare the feature
in the observed STIS spectrum of TW Hya and in the STIS spectrum
smoothed to the resolution of the ACS spectra.
While the \h2 lines are no longer observable in the
smoothed spectrum, the feature at 1600 {\AA} is.

\begin{figure}
\plotone{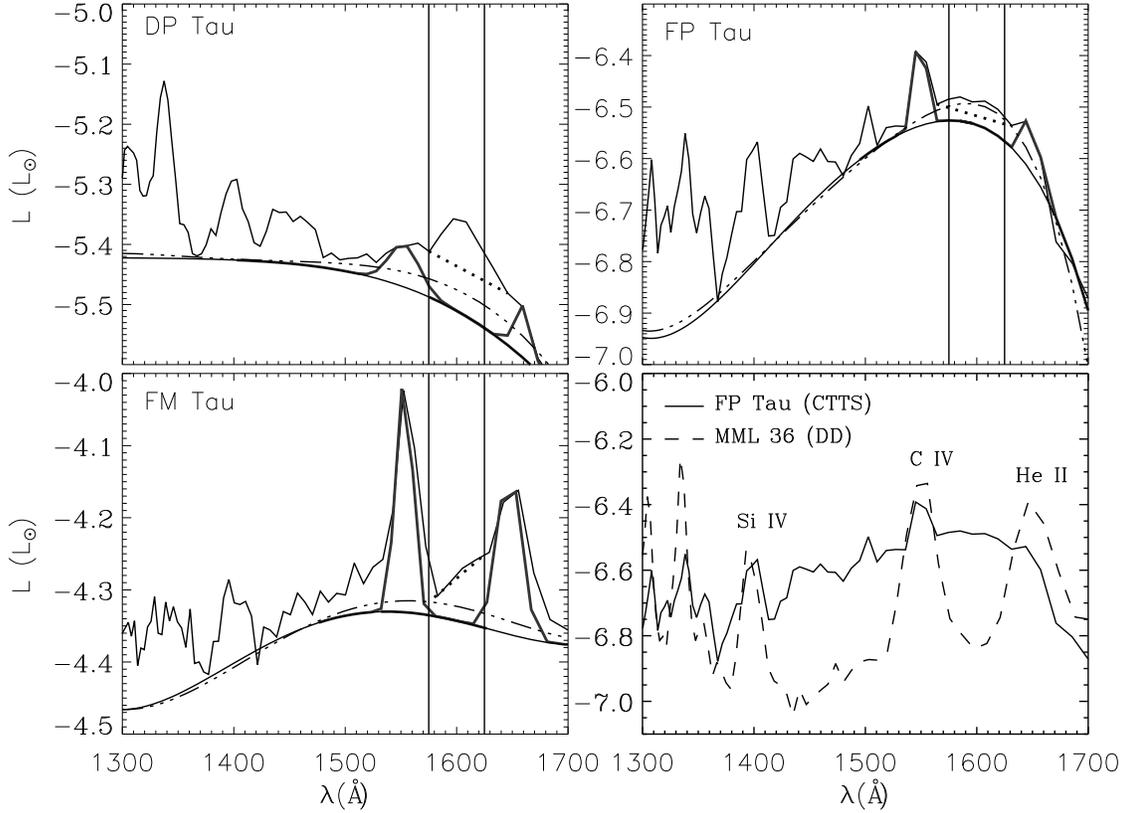}
\caption{\h2 measurements for ACS sources.  The first three panels show ACS sources and the location of the subtracted continua, shown as the solid, dashed and dot-dashed lines.  Also plotted are the He II and C IV emission lines as the thick solid line.  The final panel compares accreting and non-accreting sources with the same luminosity.  An excess in FP Tau is observed at 1600 {\AA}, which is due to electron impact \h2 emission, and also between the S IV and C IV lines, which is likely due to  blended electron impact and Ly$\alpha$ fluorescent lines.
}
\label{four}
\end{figure}

In addition to electron impact \h2 emission, the flux at 1600 {\AA} has contributions from accretion shock emission and Ly$\alpha$ fluorescent lines (Ingleby et al. 2009).  Attempting to isolate an indicator that is due to electron impact \h2 emission, we measured the flux between 1575 and 1625 {\AA} and subtracted from it the continuum and the contribution from nearby strong lines (He II 1640 {\AA} and C IV 1550 {\AA}).  Since it is unclear how strong the emission from additional sources is at 1600 {\AA}, we calculated the continuum in
three ways.  First, by joining the troughs in the spectrum on either side of the 1600 {\AA} feature; second, by fitting a 5th order polynomial to the entire FUV spectrum; third, by adopting a continuum which assumes that the rise in the spectrum at 1600 {\AA} is due entirely to electron impact \h2 emission.  Figure \ref{smtw} shows the location of the subtracted continuum for each method in TW Hya, and Figure \ref{four} shows examples of the measurements for three ACS targets.  These three methods for measuring the \h2 feature luminosity were used to estimate the errors.  Comparing the TW Hya spectra at both resolutions indicates that the feature luminosity decreases by $\sim$2 in the low resolution spectrum because some of the flux is blended into the continuum.  This error is small compared to the uncertainty in the continuum location.

\begin{figure}
\plotone{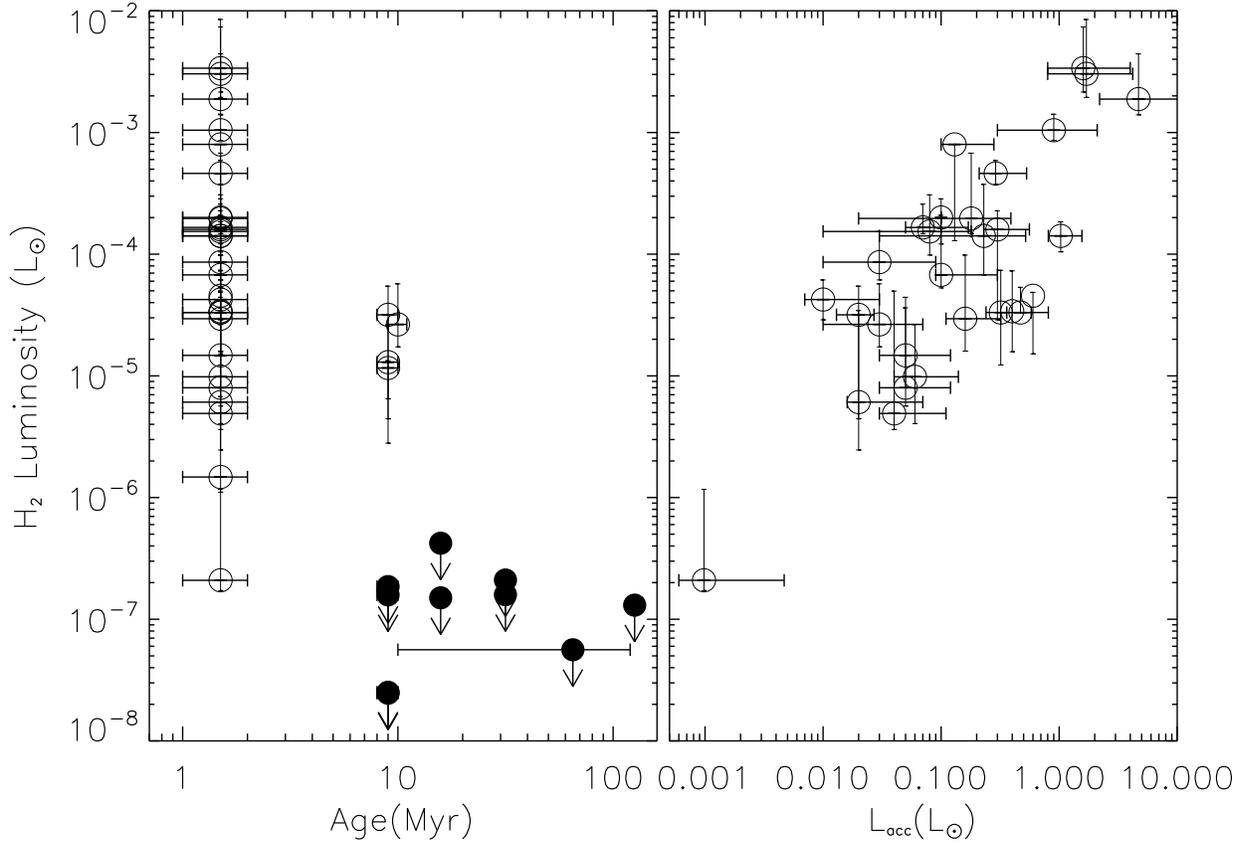}
\caption{Left: Luminosity of the \h2  vs. age.  Filled circles represent WTTS and open circles represent CTTS.  For the WTTS we show only an upper limit on the luminosity of the \h2.  Right:  \h2 luminosity vs. $L_{acc}$.  The \h2 luminosity is observed to increase with $L_{acc}$.  Errors on $L_{acc}$ are calculated using the scatter in the correlation with $L_U$ presented in \cite{gullbring98}. 
}
\label{h2age}
\end{figure}

Using these procedures, we measured the luminosity of the 1600 {\AA}
feature in both the ACS spectra and the STIS spectra smoothed to the
resolution of ACS; the feature luminosities are given in Table 1. 
For the WTTS/DD, we find that the \h2 feature is not observable and the values presented in Table 1 are upper limits based on the rms fluctuations from 1575 to 1625 {\AA}.  We thus find that
the \h2 feature shows only in the accreting sources.  This is not an age effect; our sample includes CTTS
and WTTS of similar age at $\sim$10 Myr (left panel
of Figure \ref{h2age}) but only the accreting sources show the \h2
feature.  Moreover, we find a clear correlation of the strength of the
feature with $L_{acc}$ in the CTTS (right panel of
Figure \ref{h2age}), with a Pearson correlation coefficient of 0.68,
indicating that the \h2 emission depends on the accretion properties
of the source and not on the age.  A similar result was found in
\cite{carmona07}, where the probability of detecting near-IR \h2 lines
was greater in sources with higher accretion rates.

\section{Discussion}
\label{disc}

Free electrons are required for
the process of electron excitation to be effective (\S 1).  Since, in turn,  high energy radiation fields are necessary to produce fast
electrons, the absence of \h2 emission in the WTTS/DD could in
principle be due to a low level of X-ray or EUV emission in these
objects relative to the CTTS.  However, \cite{telleschi07} found that
there is little difference between the X-ray luminosities of CTTS and
WTTS in their X-ray survey of pre-main sequence objects in Taurus.
Even though there is a soft X-ray excess created in the accretion
shock region of CTTS (G{\"u}nther et al. 2007, and references therein),
it does not significantly increase the X-ray production in most young
stars \citep{telleschi07}.  
Similarly, \cite{kastner97}
showed that CTTS and WTTS in the 10 Myr TW Hya Association have
similar X-ray luminosities.  Moreover, the X-ray luminosity does not decrease significantly
over the first 100 Myr of low mass stars \citep{briceno97,kastner97}, so
the CTTS
and WTTS/DD in our sample should have comparable X-ray luminosities.

The EUV radiation field, including emission from approximately
100 to 1000 {\AA}, is also responsible for the ionization of heavy
atoms, contributing to the population of free electrons available to
excite an \h2 molecule.  The EUV is difficult to investigate because
the radiation is extremely extincted by interstellar hydrogen.
\cite{alexander05} find that the EUV flux level does not change
in the first $\sim$10 Myr, from studies of the ratio He II 1640/CIV 1550 {\AA}.  If we assume that the FUV level is an
indicator of the strength of the EUV emission, we come to similar conclusions.  Figure \ref{four} shows one CTTS and one DD that have the same FUV luminosity, so one would expect a strong enough EUV radiation field in both sources to create the free electrons needed to excite \h2 if it were present.  However,
the excess emission at 1600 {\AA} is clearly seen in the CTTS (FP Tau) and absent in the DD (MML 36).

Since the high energy radiation fields in both CTTS and WTTS/DD are
comparable in strength, the most likely explanation for the lack of
\h2 emission in WTTS/DD is that there is essentially no gas in their
inner disks.  Given the close relationship between the \h2 feature
strength and $L_{acc}$ shown in Figure \ref{h2age}, our results
suggest that \h2 gas dissipates in timescales consistent with the
cessation of accretion; when the gas is dissipated in the inner disk,
there is no material left to accrete.

We use the observations to make a rough estimate of the column
density of \h2 being collisionally excited.  We assume that the \h2 is emitted in an optically thin region of the disk
with area $A$ and thickness $z$. The emitted luminosity per unit volume is
${\cal E}_{\lambda}=h\nu\sigma_{\lambda} v\chi_en_{H_2}^2$, where $h\nu$ is the energy of
the emitted photon, $\sigma_{\lambda}$ the \h2 cross section, $v$
the impacting electron velocity, $n_e$ the electron number
density, $\chi_e$ the electron fraction, and $n_{H_2}$ the
number density of \h2.  The expected flux at
1600 {\AA} due to electron impact excitation is then 
\be
\label{h2}
F_{1600}=
\frac{h \nu \sigma_{1600} v \chi_e \Sigma^2 R^2}{16 m_H z d^2}.  
\en
where $\Sigma$ is the surface density of \h2 excited by electron
impacts, $m_H$ the mass of hydrogen,
$R$ the radius of the emitting region,
and $d$ is the distance.  In
Ingleby et al. (2009) we find that the electron excitation model that
provides the best fit 
to the 1600 {\AA} feature of our sample of CTTS
with STIS spectra is characterized by a temperature $T \sim$ 5000 K
and an electron energy of $\sim$12 eV. For these values, $\sigma_{1600} =
10^{-20} {\rm cm^2 {\AA}^{-1}}$ \citep{abgrall97}. According to the
thermal models of \cite[][M08]{meijerink08}, gas reaches $T \sim 5000$K
within 1 AU of the star, which is consistent with the upper limit to
the extension of the \h2 emitting region set by the STIS resolution in
the case of TW Hya \citep{herczeg02}. We further
assume that most electrons are capable of exciting \h2 and adopt
$\chi_e=5\times10^{-3}$, as well as $R\sim$1 AU and $z\sim$0.1 AU (M08).  Using these numbers, and assuming that all the flux at 1600 {\AA} is
due to electron impact excitation, we get the estimates of $\Sigma$ in
Table 1, which for CTTS are consistent with predicted
formation in the
uppermost levels of the disk (M08).

A similar estimate can be made for the column density of electron
excited \h2 in the WTTS/DD in our sample, which
have some
dust remaining at larger radii but no detected IR \h2 lines
\citep{carpenter09,carpenter08,hillenbrand08,verrier08,chen05,low05}.
These estimates are given in Table \ref{tabprop}.  We used the flux of MML 36, which is the
WTTS/DD with the highest flux at 1600 {\AA} in our sample, to estimate
the mass of \h2 inside $\sim$1 AU; we found that there must be less than
$10^{-7}$ earth masses, $10^{-7}$\% of the MMSN,
lower than the 0.01\% of the MMSN estimated by \cite{pascucci06}.  This has important implications
for the formation of terrestrial planets, especially if gas is needed
to circularize orbits \citep{agnor02}.  \cite{kominami02} theorize
that at least 0.01\% of the MMSN must be present
during the formation of proto-planets, which form around 10 Myr
according to simulations by \cite{kenyon06}.  Our column density estimates
indicate that the amount
of \h2 gas present in WTTS/DD with ages of 10-100 Myr is too small to
circularize the orbits of the terrestrial planets being formed at that
time.  Our results support the conclusion by \cite{pascucci06} that
there must be an additional source responsible for damping
eccentricities, one possibility being dynamical friction with
remaining planetesimals.  Another possibility is that other species of
gas exist after the \h2 has been depleted, for example, C and O have
been detected around the 10 Myr debris disk $\beta$ Pic
\citep{fernandez06,roberge06}.  C and O do not feel strong
radiation pressure due to the low FUV flux in WTTS and therefore may
remain after the \h2 has been depleted \citep{roberge06}.

\section{Acknowledgments}
We thank Al Glassgold for discussions clarifying the ionization
mechanisms in the disk.  This work was supported by NASA through
grants GO-08317, GO-09081, GO-9374, GO-10810 and GO-10840 from the Space Telescope Science
Institute.  This material is also based upon work supported by the
National Science Foundation under Grant No. 0707777 to EAB.

\begin{deluxetable}{lcccccccc}
\tablewidth{0pt}
\tablecaption{ Sources
\label{tabprop}}
\tablehead{
\colhead{Object} & \colhead{Spectral Type}   & \colhead{$L$} &  \colhead{$A_V$} & \colhead{Age}&\colhead{$L_{acc}$} & \colhead{\h2 Feature} & \colhead{$\Sigma$}  \\
\colhead{ }       &\colhead{ }       & \colhead{$\lsun$} & \colhead{ mag}& \colhead{Myr} & \colhead{$\lsun$} &\colhead{ $10^{-5} \lsun$} & \colhead{$10^{-6}$ g cm$^{-2}$}  \\}
\startdata
ACS CTTS\\
AA Tau & M0  & 1.1 & 1.4  & 1  &  0.13$\pm_{.03}^{.15}$    &  79.9$\pm_{67}^{0}$ & $>$ 49.3\\
CI Tau  & K6   & 1.3  & 2.1  & 1  & 0.47$\pm_{.11}^{.34}$     &  3.3$\pm_{0}^{2.0}$   & $>$ 9.9\\
DE Tau & M1   & 1.2 & 1.1  & 1  & 0.16$\pm_{.05}^{.16}$    &  2.9$\pm_{1.4}^{6.9}$   & $>$ 36.3 \\
DL Tau & K7    & 1.0 & 1.6  & 1  & 0.32$\pm_{.08}^{.26}$    &  3.3$\pm_{2.1}^{4.1}$   & $>$ 22.5\\
DN Tau & M0   & 1.2 & 0.8  & 1  & 0.04$\pm_{.01}^{.07}$    &  0.49$\pm_{.13}^{4.5}$ & $>$ 18.4\\ 
DO Tau & M0  & 1.4 & 2.4 & 1  & 0.29$\pm_{.08}^{.24}$     &   46.1$\pm_{8.6}^{13}$ & $>$ 84.6\\
DP Tau & M0   & 0.2 & 0.5 & 1  & 0.01$\pm_{.003}^{.02}$     &  4.2$\pm_{1.4}^{1.9}$    & $>$ 17.6\\
DR Tau & K7  & 1.7 & 1.0  & 1 & 1.03$\pm_{.22}^{.53}$     &  14.1$\pm_{3.7}^{4.3}$  & $>$ 43.2\\
FM Tau & M0  & 0.5 & 1.9 & 1 & 0.30$\pm_{.07}^{.26}$       &  16.0$\pm_{13}^{6.8}$  & $>$ 61.7\\
FP Tau &  M3  & 0.4 & 0.1  & 1 & 0.001$\pm_{.0004}^{.004}$ &  0.021$\pm_{.004}^{.10}$ & $>$ 4.9\\
GK Tau & M0  & 1.4 & 1.1 & 1  & 0.06$\pm_{.02}^{.08}$      &  0.98$\pm_{.58}^{1.7}$   & $>$ 18.1\\
HN Tau A$^{\ast}$ & K5  & 0.2 & 1.2  & 1  & 0.07$\pm_{.02}^{.10}$     &  16.6$\pm_{1.8}^{9.2}$   & $>$ 38.8\\
HN Tau B$^{\ast}$& M4  & 0.03&0.9  & 1  & --         &0.15$\pm_{.04}^{.53}$      & $>$ 5.8\\
IP Tau   & M0  & 0.7 & 0.9  & 1  & 0.02 $\pm_{.004}^{.05}$    &  0.61$\pm_{0.4}^{2.8}$   & $>$ 15.0\\
UZ Tau A$^{\ast}$ & M1  & 0.3 & 0.5  &  1 & 0.02$\pm_{.02}^{.07}$       &   0.80$\pm_{0}^{2.8}$   & $>$ 14.3\\
UZ Tau B$^{\ast}$ & M2  & 0.3  & 1.0  &  1 & 0.02$\pm_{.02}^{.07}$       & 1.5$\pm_{.91}^{2.9}$      & $>$ 8.5\\
CVSO 206&K6& 0.2 & 0.2  & 9  & --            & 1.2$\pm_{.51}^{.17}$       & $>$ 13.9\\
CVSO 35 &K7&0.7&0.7  & 9   & 0.02$\pm_{.01}^{.01}$       &3.2$\pm_{2.7}^{2.3}$        & $>$ 16.6 \\
CVSO 224$^{\dagger}$&M3&0.1&0.5  &9    &--            &--             &--\\
OB1a 1630&M2&1.0&0.0   &9    &--            &1.3$\pm_{1.0}^{1.3}$         &$>$ 13.3\\
\hline
STIS CTTS\\
BP Tau   &  K7  & 1.3 & 1.0  & 1  & 0.23$\pm_{.20}^{.29}$  &  14.1$\pm_{7.4}^{23}$ & $>$41.6\\
DM Tau  &  M1  & 0.3 & 0.6  & 1  & 0.08$\pm_{.07}^{.10}$  &  15.4$\pm_{5.5}^{15}$ & $>$ 39.5\\
GM Aur   &  K3  & 1.2 & 1.1  & 1  & 0.18$\pm_{.16}^{.21}$  &  19.7$\pm_{4.9}^{48}$ & $>$ 48.7\\
LkCa 15 &  K5  & 1.0 & 1.0  & 1  & 0.03$\pm_{.02}^{.06}$  &  8.6$\pm_{2.5}^{6.8}$ & $>$ 26.4\\
RY Tau   &  G1  & 9.6 & 2.2  & 1  & 1.6$\pm_{.80}^{2.4}$   &  338.0$\pm_{120}^{400}$  & $>$ 148.4\\
SU Aur   &  G1  & 7.8 & 0.9  & 1  & 0.10$\pm_{.01}^{.20}$  &  6.8$\pm_{1.5}^{14}$ & $>$ 30.0\\
T Tau      &  G6  & 7.8 & 1.8  & 1  & 0.90$\pm_{.60}^{1.2}$  &  104.5$\pm_{18}^{37}$ & $>$ 103.9\\
CO Ori    &  G0  & 22.3 & 2.0  & 1  & 1.7$\pm_{.90}^{2.5}$  &  303.5$\pm_{110}^{550}$   & $>$ 149.0\\
EZ Ori     &  G3  & 5.9 & 0.6  & 1  & 0.10$\pm_{0}^0$  &  20.0$\pm_{7.9}^{8.5}$ & $>$ 41.1\\
GW Ori   &  G0  & 61.8 & 1.3  & 1  & 4.7$\pm_{2.5}^{6.9}$  &  188.2$\pm_{49}^{250}$   & $>$ 178.8\\
P2441    &  F9   & 11.5&  0.4 &  1  & 0.4$\pm_{.20}^{.60}$  &  3.4$\pm_{1.8}^{3.9}$        & $>$ 29.0\\
V1044 Ori&G2  & 6.7 &   0.4&   1  &  0.6$\pm_{.30}^{.90}$  &  4.6$\pm_{3.0}^{.30}$       & $>$ 37.6\\
TW Hya  &  K7  & 0.3 & 0     & 10  & 0.03$\pm_{.02}^{.04}$  &  2.6$\pm_{.92}^{3.1}$ & $>$ 43.9\\
\hline
ACS WTTS\\
HD 12039    &  G4   & -- & 0      & 31.6  & 0  &   $<$.021 & $<$ 2.2\\
HD 202917  &  G5   & 0.7 & 0  & 31.6  & 0  &   $<$.016    &$<$ 2.8\\
HD 61005    &  G8   & 0.6 & 0      & 125.9 & 0  &   $<$.013 & $<$ 1.3 \\
HD 92945    &  K2   &  --& 0   & 20 - 150  & 0  &   $<$.006     & $<$ 0.93\\
HD 98800    &  K5   & 0.6  & 0   & 10.0  & 0   &   $<$.019    & $<$ 0.84\\
MML 28        &  K2   & -- & 0.1   & 15.8  & 0  &   $<$.015  & $<$ 2.4\\
MML 36        &  K5   &--  & 0.3   & 15.8  & 0  &   $<$.042 & $<$ 3.3\\
TWA 7           &  M1  & 0.31  & 0      & 10.0  & 0  &   $<$.003   &$<$ 1.0\\
TWA 13A      &  M1  & 0.18 & 0      & 10.0  & 0  &   $<$.016   & $<$ 1.5\\
TWA 13B      &  M1  & 0.17 & 0      & 10.0  & 0  &   $<$.002   & $<$ 1.5\\
\enddata
\tablecomments{\\
$^{\ast}$ Stellar properties for binaries are from \cite{white01}.  UZ Tau A and B are themselves binaries; UZ Tau A is a spectroscopic binary and UZ Tau B is a binary system \citep{white01} but is unresolved by ACS/SBC.\\
$^{\dagger}$ CVSO 224 is a CTTS surrounded by a transitional disk \citep{espaillat08} and has a very low $\mdot$.  The ACS/SBC spectrum of this target is noisy and while we cannot confidently quantify the \h2 emission, we do see the rise in the spectrum at 1600 {\AA} which indicates its presence.}
\end{deluxetable}

\clearpage

\clearpage

\clearpage

\clearpage

\end{document}